\documentclass[pre,twocolumn,groupedaddress,floatfix]{revtex4-1}
\usepackage{amssymb,amsmath}
\usepackage{graphicx}
\usepackage{subfigure}
\usepackage[english]{babel} 
\usepackage{float}
\usepackage{color}

\usepackage{lipsum}
\begin{document}
\newcommand{\be}{\begin{equation}}
\newcommand{\ee}{\end{equation}}
\newcommand{\rojo}[1]{\textcolor{red}{#1}}

\title{Fractional electrical impurity}

\author{Mario I. Molina}
\affiliation{Departamento de F\'{\i}sica, Facultad de Ciencias, Universidad de Chile, Casilla 653, Santiago, Chile}

\date{\today }

\begin{abstract} 
We examine the localized mode and the transmission of plane waves across a capacitive impurity of strength $\Delta$, in a 1D bi-inductive electrical transmission line where the usual discrete Laplacian is replaced by a fractional one characterized by a fractional exponent $s$. In the absence of the impurity, the plane wave dispersion is computed in closed form in terms of hypergeometric functions. It is observed that the bandwidth decreases steadily, as $s$ decreases towards zero, reaching a minimum width at $s=0$. The localized mode energy and spatial profiles are computed in close form v\`{i}a lattice Green functions. The profiles show a remnant of the staggered-unstaggered symmetry that is common in non-fractional chains. The width of the localized mode decreases with decreasing $s$, becoming completely localized at the impurity site at $s=0$. The transmission coefficient of plane waves across the impurity is qualitatively similar to its non-fractional counterpart ($s=1$), except at low $s$ values ($s\ll 1$). For a fixed exponent  $s$, the transmission decreases with increasing $\Delta$.

\end{abstract}

\maketitle

{\em Introduction}. The study of the effects of disorder on the transport properties of an excitation propagating in an otherwise discrete, periodic lattice, continues to be a topic of interest despite its long history. Examples of this include chains of atoms and optical waveguide arrays where the eigenvalues form a well-defined band with eigenvectors that are extended waves\cite{uno}. The insertion of a single impurity breaks the discrete translational invariance and causes one of the states at the band edge to detach from the band giving rise to a localized mode centered at the impurity position. The rest of the modes remain extended albeit somewhat distorted. In 1D and 2D lattices, there is always a localized bound state centered at the impurity\cite{dos,tres}, no matter how weak the strength of the impurity.

Studying a single impurity serves as an initial step toward understanding more intricate disordered systems. This is particularly relevant when dealing with systems that contain a finite fraction of impurities. In such cases, the prominent phenomenon under investigation is Anderson localization which states that the presence of disorder causes all modes to be spatially localized in 1D and 2D, while in 3D a mobility edge is created. In fact, it has been suggested that there exists an underlying connection between the problem of finding the localized mode in a system with  a single impurity, with that of computing the localization of the modes in the Anderson system\cite{cuatro,cinco}. 
Several examples of linear impurities include coupling defects, discrete networks used for routing and switching of discrete optical solitons\cite{seis}, junction defects between two optical or network arrays\cite{siete}, as well as simple models for magnetic metamaterials. In the latter case, magnetic metamaterials are modeled as periodic arrays of split-ring resonators, where magnetic energy can be trapped at impurity positions\cite{ocho}. Other related applications abound in a diverse array of themes, ranging from defect modes in photonic crystals\cite{nueve}, optical waveguide arrays\cite{diez,once,doce,trece}, superconductors\cite{catorce}, dielectric superlattices with embedded defect layers\cite{quince,dieciseis}, electron-phonon interactions\cite{dieciseis}, granular crystals in materials science\cite{diecisiete}, and electrical transmission lines. In this last case,  localized modes at capacitive impurity places have been experimentally observed in 1D\cite{dieciocho}.

For a single impurity, the usual approach is to assume an exponential shape for the localized mode profile. There is no guarantee, however, that this empirical approach will work in all cases. For instance, for problems involving boundaries, like impurities close to a surface, a more serious treatment is needed. An elegant method for dealing with impurity problems is the technique of lattice Green functions\cite{diecinueve, veinte, veintiuno}. This is the method we will follow in this work, with the added feature of fractionality. 

On the other hand, fractionality is a concept that has gained considerable interest in recent years. Its origin dates back to about 300 years ago when some mathematicians started discussing possible extensions of Newton's calculus. The main question then was to ascertain whether an integer order derivative  could be extended to non-integer values. Early work suggested that this was possible in principle and thus, these initial studies were followed by rigorous work by Euler, Laplace, Riemann, and Caputo to name a few, and elevated fractional calculus from a mere mathematical curiosity to a fully established research field.\cite{veintidos,veintitres,veinticuatro,veinticinco}. Several possible definitions for the fractional derivative have been given, each one with its advantages and disadvantages. The two most popular definitions are the Riemann-Liouville form
\be
\left({d^{}\over{d x^{s}}}\right) f(x) = {1\over{\Gamma(1-s)}} {d\over{d x}} \int_{0}^{x} {f(s)\over{(x-s)^{s}}} ds \label{1}
\ee
and the Caputo form
\be
\left({d^{s}\over{d x^{s}}}\right) f(x) = {1\over{\Gamma(1-s)}} \int_{0}^{x} {f'(s)\over{(x-s)^{s}}} ds.  \label{2}
\ee
where, $0<s<1$. This fractional formalism has found applications in several fields: 
Levy processes in quantum mechanics\cite{veintiseis}, strange kinetics\cite{veintisiete}, optics\cite{veintiocho,veintinueve,treinta,treinta y uno,treinta y dos,treinta y tres}, fluid mechanics\cite{treinta y cuatro}, quantum mechanics\cite{treinta y cinco,treinta y seis}, electrical propagation in cardiac tissue\cite{treinta y siete}, epidemics\cite{treinta y ocho} and biological invasions\cite{treinta y nueve}. 

One of the main features of fractional formalism is that it incorporates memory effects or nonlocal effects, as can be seen from the structure of Eqs.(\ref{1}) and (\ref{2}). It is interesting then, to determine the interplay between disorder and nonlocality by means of a simple classical model: An electrical  excitation  propagating along a 1D electrical chain in the presence of a single impurity with a discrete fractional Laplacian. More concretely, we will consider a 1D bi-inductive electric transmission line (Fig.1) where one of the capacitors differs from the rest. For instance, it could be filled with a different dielectric or the spacing between the capacitor plates could be different from the rest.  We will look at the behavior of the electric charge inside the capacitors and focus on the properties of the localized mode and that of the transmission of plane waves across the impurity.

{\em The model}. Let us consider an electrical lattice composed of a one-dimensional array of $LC$ circuits coupled inductively, containing a single capacitive impurity (Fig.1). $L_{1}$ and $L_{2}$ are the inductances and $C_{n} = C + (C_{0}-C)\;\delta_{n 0}$, where $C$ is the capacitance in vacuum and $C_{0}$ is the impurity capacity obtained by inserting a dielectric between the plates of the capacitor at $n=0$, or by changing the spacing between its plates. The electrical charge $Q_{n}$ on the $n$th capacitor is given by $Q_{n} = C_{n} U_{n}$. After using Kirchhoff's law, the equations for the voltages are
\be
{d^2 Q_{n}\over{d t^2}} = {1\over{L_{1}}}(U_{n+1}-2 U_{n}+U_{n-1}) - {1\over{L_{2}}}U_{n}.
\ee
After inserting the expression for $Q_{n}$ in terms of $U_{n}$ and after introducing dimensionless variables, we obtain
\be
(1+\Delta\;\delta_{n 0}) {d^2\over{d \zeta^2}} V_{n} = (V_{n+1}-2 V_{n}+V_{n-1})-\omega^2 \ V_{n},\label{eq4}
\ee
where $\Delta=(C_{0}/C)-1$, which implies a physical range $-1<\Delta<\infty$,  $\omega^2=(\omega_{2}/\omega_{1})^2$, where $\omega_{1}=1/\sqrt{L_{1} C_{0}}$, $\omega_{2}=1/\sqrt{L_{2} C_{0}}$ are the resonant frequencies, $\zeta=\omega_{1} t$ is the dimensionless time, and $V_{n}=U_{n}/U_{c}$, where $U_{c}$ is a characteristic voltage.
\begin{figure}[t]
 \includegraphics[scale=0.15]{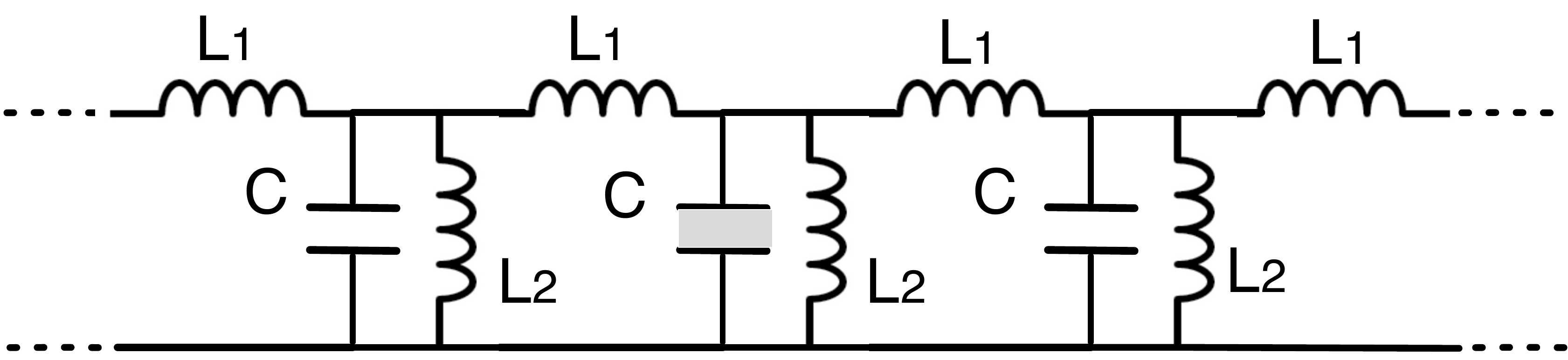}
  \caption{Infinite (top) and semi-infinite (bottom) 
  bi-inductive electrical lattice containing a capacitive impurity (after ref.\cite{circuit}).}  \label{fig1}
\end{figure}

We identify the first term on the RHS of Eq.(\ref{eq4}), as the discrete Laplacian $\Delta_{n} V_{n}$. Thus, we can write
\be
(1+\Delta\;\delta_{n 0}) {d^2\over{d \zeta^2}} V_{n} = \Delta_{n} V_{n}-\omega^2 \ V_{n},\label{eq5}
\ee
We proceed now to promote this discrete one-dimensional Laplacian to its fractional form, by using results by Roncal et al.\cite{discrete laplacian} in which an expression is obtained for the $s$-th power of the discrete Laplacian:
\be
(-\Delta_{n}^{s}) V_{n} = \sum_{m\neq n} K^{s}(n-m) (V_{n}-V_{m})
\ee
where
\be
K^{s}(m) = {4^{s} \Gamma(s + (1/2))\over{\sqrt{\pi} |\Gamma(-s)|}} {\Gamma(|m|-s)\over{\Gamma(|m|+1+s})},
\ee
and $\Gamma(x)$ is the Gamma function. 
Thus, the main equation can be written as 
\begin{eqnarray}
& & (1+\Delta\;\delta_{n 0}) {d^2\over{d \zeta^2}} V_{n}+ \sum_{m\neq n} K^{s}(n-m) (V_{n}-V_{m})\nonumber\\
& &\hspace{1cm}+ \omega^2 \ V_{n}=0,\label{eq10}
\end{eqnarray}
We can see that the effect of a fractional discrete laplacian is to introduce nonlocal interactions via a symmetric kernel $K^{s}(n-m)$. Using the relation $\Gamma(n+s)=\Gamma(n) n^s$, we can obtain the asymptotic decay $K^{s}(m)\rightarrow 1/|m|^{1 + 2 s}$ at long distances, that is, a power-law decrease of the coupling with distance instead of the more familiar exponential decay.
\begin{figure}[t]
 \includegraphics[scale=0.25]{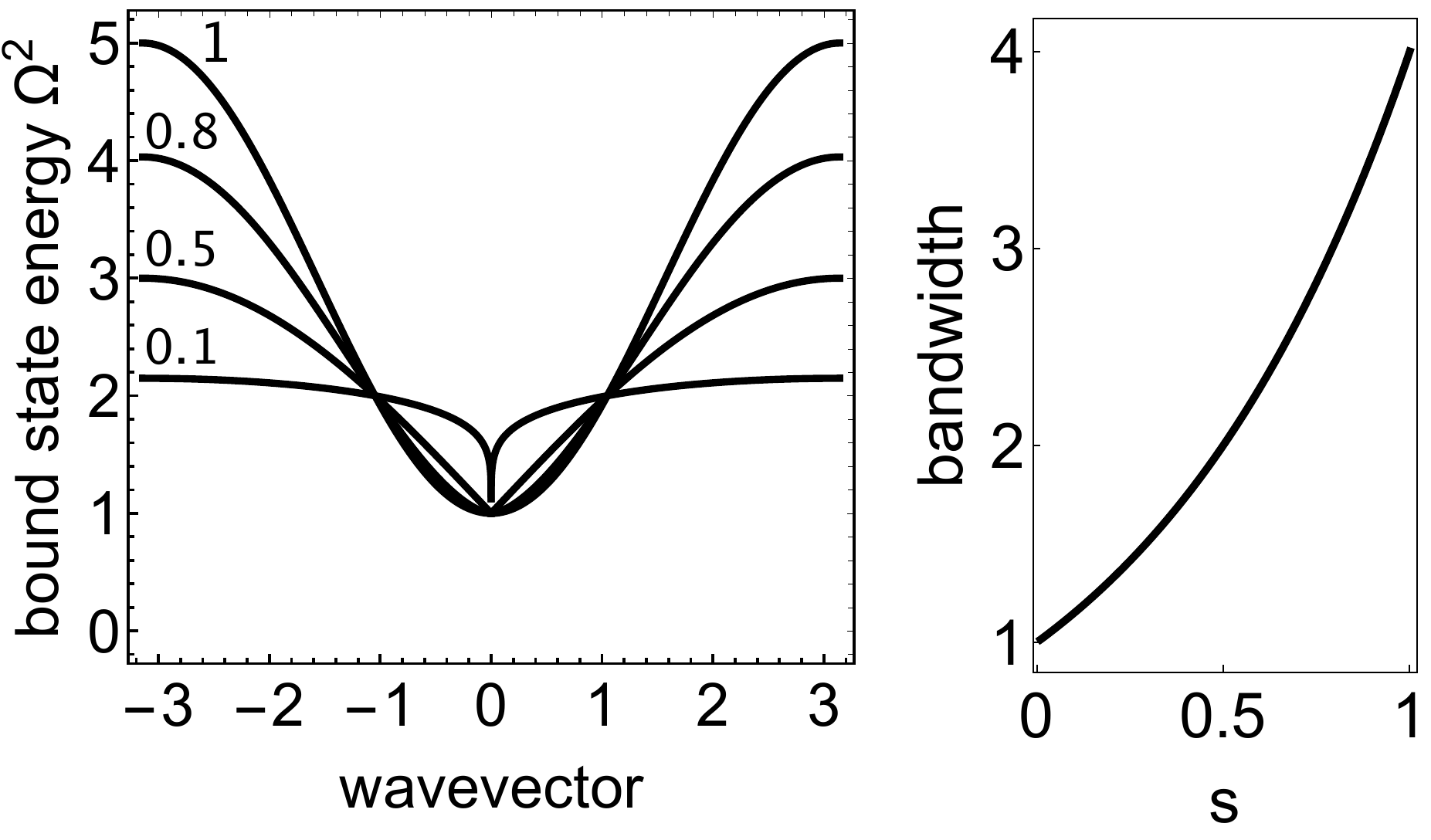}
  \caption{Left: Dispersion of electro-inductive waves. The number on each curve denote the value of the fractional exponent $s$. Right: Bandwidth vs fractional exponent. ($\omega^2=1$)}  \label{fig2}
\end{figure}

In the absence of the impurity, the stationary-state equation can be written as (after setting $V_{n}(\zeta)=V_{n} \cos(\Omega \zeta + \phi)$):
\be
-\Omega^2\;V_{n}+ \sum_{m\neq n} K^{s}(n-m) (V_{n}-V_{m})+\omega^2 \ V_{n}=0.\label{eq9}
\ee
We now look for the dispersion relation of plane waves $V_{n} = A\ e^{i k n}$. We obtain
\be
\Omega^2(k) = \omega^2+ 4 \sum_{m=1}^{\infty} K^{s}(m) \sin^2(k\;m /2)\label{eq10}.
\ee
This can be expressed in closed form as
\begin{eqnarray}
\Omega^2(k)&=&\omega^2+{e^{-i k}\over{\pi \Gamma(1+s)}}( -\pi s \Gamma(1+2 s)(\  _{2}F^{+}_{1}(k)+\nonumber\\
\lefteqn{+e^{2 i k}  {_{2}F^{-}_{1}}(k)\ )+2 e^{i k} \Gamma(1-s)\Gamma(2 s)\sin(\pi s) )}.  \label{eq11}
\end{eqnarray}
where, $F^{+}(k)={_{2}F_{1}}(1,1-s,2+s,e^{-i k})$, $F^{-}(k)=_{2}F_{1}(1,1-s,2+s,e^{i k})$, and $_{2}F_{1}$ is the regularized hypergeometric function.
Figure 2 shows
$\Omega^2(k)$ vs $k$ for several values of the fractional exponent  $s$. Figure 2 also shows the bandwidth vs the fractional exponent $s$. The bandwidth is defined as $\Delta B(s)=|\Omega^2(\pi)-\Omega^2(0)|$. By taking the appropriate limits in Eq. (\ref{eq11}), we have $\Delta B(s\rightarrow 1)=4$ and $\Delta B(s\rightarrow 0)=\omega^2$. Moreover, $\Delta B(0)=\omega^2$, independent of the fractional exponent $s$.

{\em Green functions.}\ The stationary equations (\ref{eq9}) can be written as
\begin{eqnarray}
\lefteqn{
(\Omega^2(k)-\omega^2)\;V_{n}+ \sum_{m\neq n} K^{s}(n-m)V_{m}+}\nonumber\\
& &-\left(\sum_{m\neq n} K^{s}(n-m)\right) V_{n}=0,
\end{eqnarray}
that is,
\be
\lambda(k)\; V_{n} = \epsilon_{n} V_{n} + \sum_{m} V_{n m} V_{m}\label{eq12}
\ee
where,
$\lambda(k)=\Omega^2(k)-\omega^2$, $\epsilon_{n}=\sum_{m} K^{s}(n-m)$, $V_{n m}=K^{s}(n-m)$.
Equation (\ref{eq12}) is formally identical to the stationary 
\begin{figure}[t]
 \includegraphics[scale=0.19]{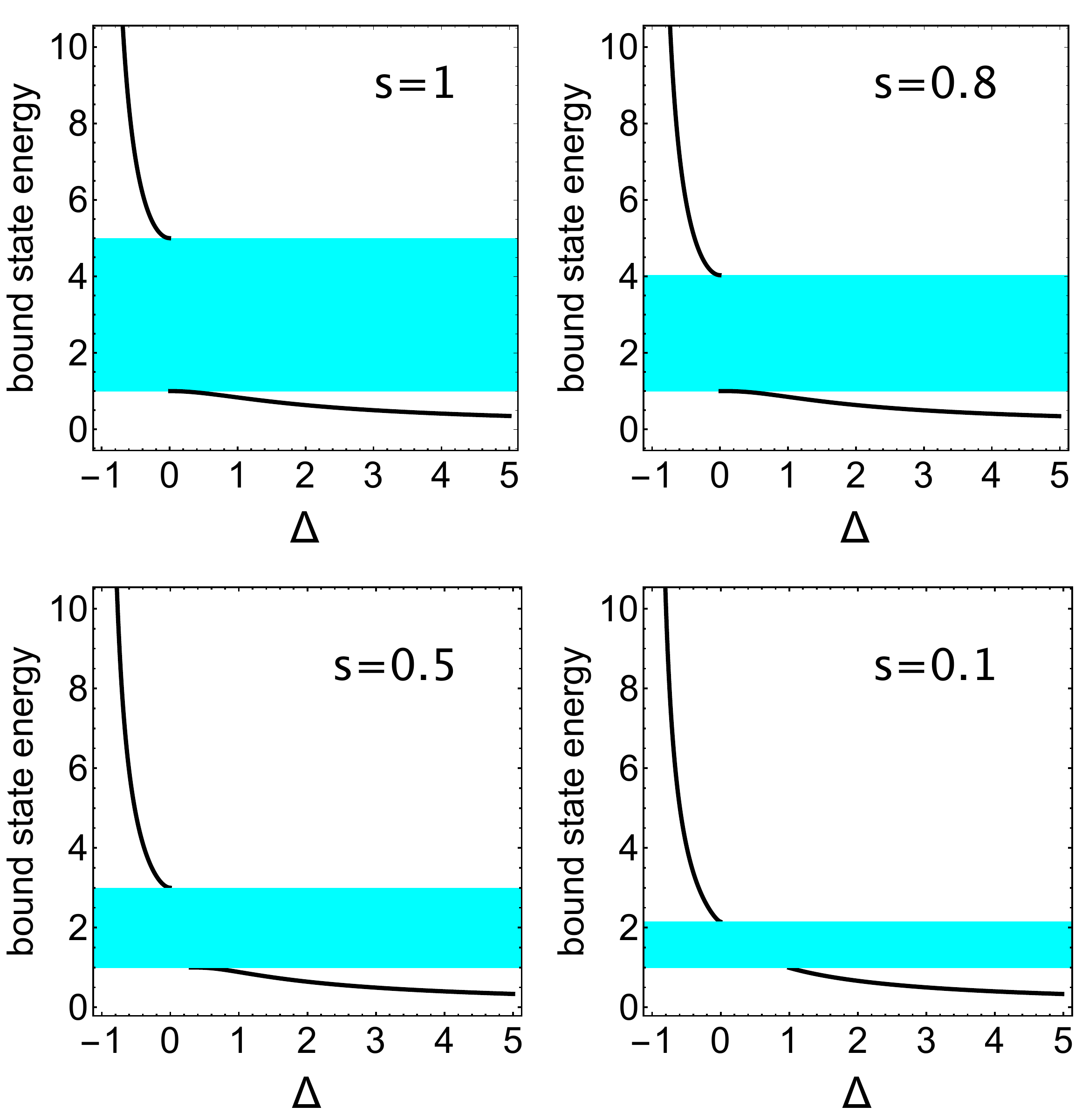}
  \caption{Bound state energy of localized mode vs impurity strength, for several values of the fractional exponent.}  \label{fig3}
\end{figure}
equation for an electron in a tight-binding model. In that case, the effective Hamiltonian is $H = H_{0} + H_{1}$ with
\be 
H_{0}=\sum_{n} \epsilon_{n}|n\rangle \langle n| + \sum_{n,m} V_{n m} |n\rangle\langle m|
\ee
\be
H_{1} = -\Delta \Omega^2(k) |0\rangle\langle 0|\label{H1}
\ee 
where the impurity has been placed at $n=0$ and Dirac's notation has been used. Hamiltonian $H$ represents a free particle propagating in the presence of a single impurity. 

The Green function is defined as
\be
G(z) = {1\over{z-H}}\label{G}
\ee
while the unperturbed Green function is given by $G^{(0)}=1/(z-H_{0})$. In an explicit form it can be written as
\be
G_{n m}^{(0)}(z)={1\over{2 \pi}} \int_{-\pi}^{\pi} {e^{i k (n-m)} dk\over{z-\lambda(k)}}\label{G0}
\ee
where $n$ and $m$ are lattice positions, and where we have used the notation $G_{m n}^{(0)} = \langle m|G^{(0)}|n \rangle$. Treating formally $H_{1}$ as a perturbation, we expand $G(z)$ as
\be
G(z) = G^{(0)} + G^{(0)}\ H_{1}\  G^{(0)} + G^{(0)}\ H_{1}\ G^{(0)}\ H_{1}\  G^{(0)}+\cdots
\ee
After inserting (\ref{H1}) for $H_{1}$, and after resuming the perturbative series to all orders, we obtain
\be
G(z) = G^{(0)} - \Delta \Omega^2 {G^{(0)}|0\rangle\  \langle 0| G^{(0)} \over{1 + \Delta \Omega^2 \ G_{0 0}^{(0)}}}.
\ee
\begin{figure}[t]
 \includegraphics[scale=0.29]{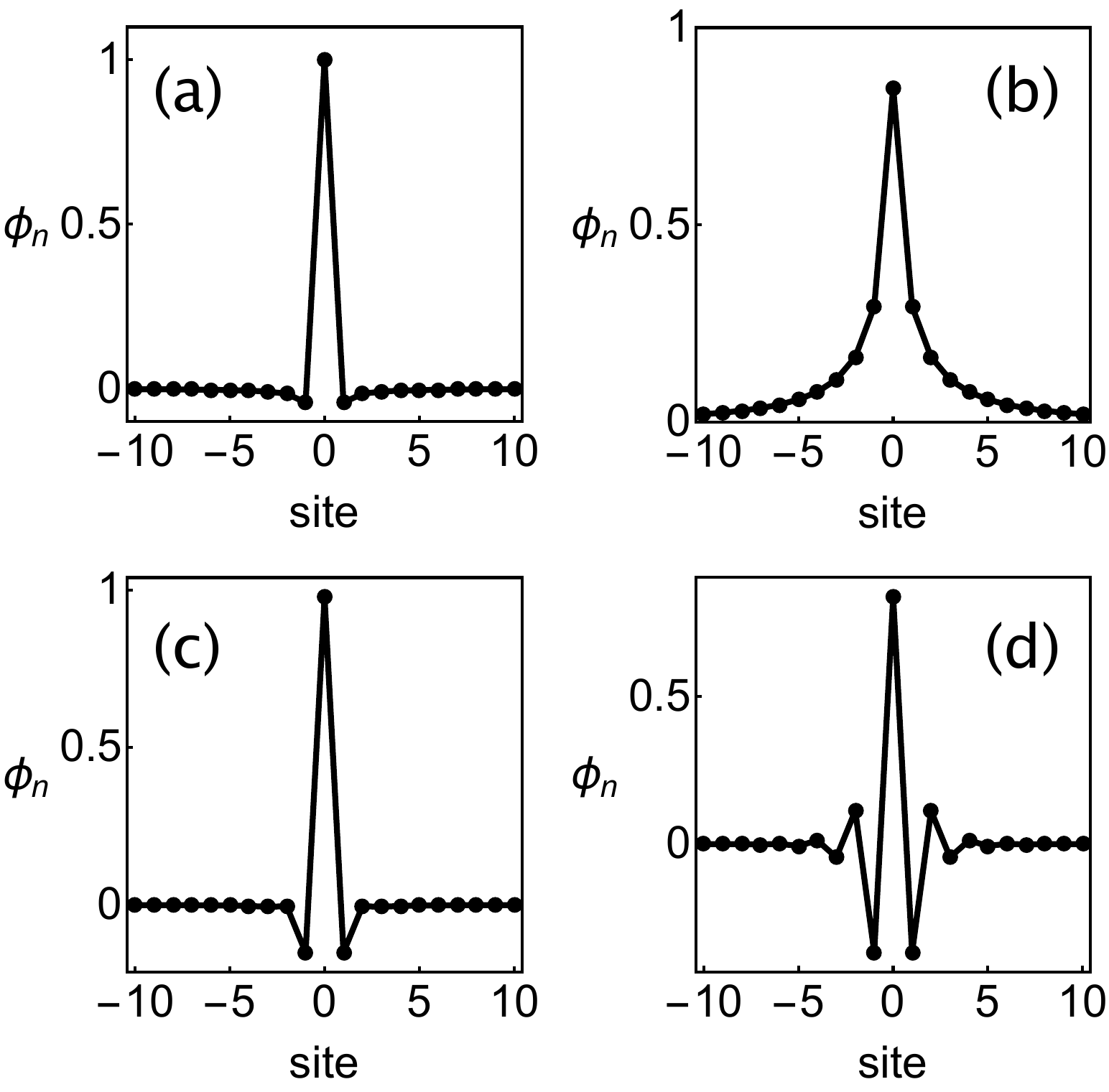}
  \caption{Examples of localized mode profiles  
  (a) $s=0.1, \Delta=-0.504, \Omega_{b}^2=4.709$,  
  (b) $s=0.5, \Delta=2.03, \Omega_{b}^2=0.636$, 
  (c) $s=0.5, \Delta=-0.511,\Omega_{b}^2=4.920$, 
  (d) $s=0.5, \Delta=-0.202, \Omega_{b}^2=3.288$. In all cases, $\omega^2=1$ }  \label{fig4}
\end{figure}
 According to the general theory\cite{diecinueve}, the energy $z_{b}$ of the bound state is given by the poles of $G_{0 0}(z)$ 
\be
1 + \Delta\Omega_{b}^2\ G_{0 0}^{(0)}(z_{b})=0,\label{eq20}
\ee 
where $z_{b}=\Omega_{b}^2-\omega^2$. 
On the other hand, the square of the mode amplitude at site $n$ is given by the residue of $G_{n m}(z)$ at the pole
\be
|\phi_{n}|^2 = -{{G_{n 0}^{(0)}}^2 (z_{b})\over{G_{0 0}'^{(0)}(z_{b})}}.
\label{phin2}
\ee

Figure 3 shows the energy of the localized mode vs the impurity strength, extracted from Eq.(\ref{eq20}).
As usual, the bound state energy lies outside the band whose width decreases as the fractional exponent 
$s$ decreases. It must be remembered that the physical range of $\Delta$ lies between $\Delta=-1$ (when $C_{0}/C \rightarrow 0$), and $\Delta \rightarrow \infty$ (when $C_{0}/C \rightarrow \infty$). In the first case, the spacing between the capacitor plates diverges, reducing its capacitance to zero, while in the second case, the spacing decreases to zero, producing a divergent capacitance. Now, a feature that is hard to appreciate in Fig.3 is the fact that, besides an energy gap in the bound state energy, there is also an `impurity gap', where a minimum value of $\Delta$ is necessary to create  the localized mode. This feature is more prominent at small fractional exponent values. For instance, at $s=0.1$ and positive $\Delta$ values, one needs $\Delta>1$ to have a bound state. For $s=0.5$, one needs $\Delta > 0.3$. At $s=1^{-}$ there is no impurity gap. On the contrary, for negative impurity strengths, there is no restriction on the magnitude of $\Delta$. 

In Fig.4 we show some examples of impurity bound state profiles for some different parameter values. While in some cases, $\phi_{n}$ decreases away monotonically  from 
\begin{figure}[t]
 \includegraphics[scale=0.19]{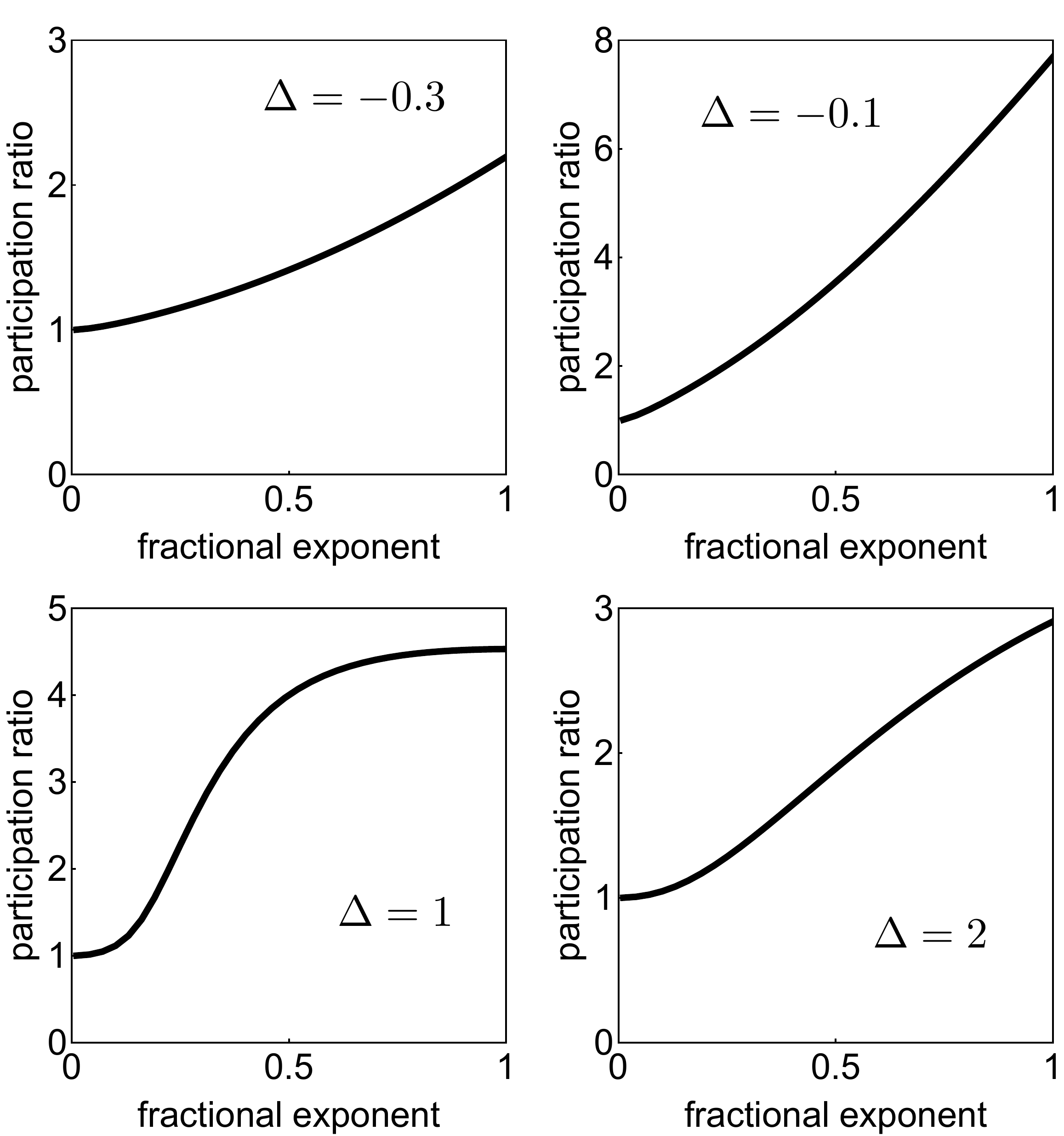}
  \caption{Participation ratio of the localized mode vs fractional exponent, for several impurity strength values.}  \label{fig5}
\end{figure}
$n=0$, in other cases we notice something like a tendency for a phase jump in $\phi_{n}$ on neighboring sites. This is a residual effect from the staggered-unstaggered symmetry obeyed in the non-fractional case ($s=1$). In that case, the equations are invariant under the transformations $\Delta\rightarrow -\Delta, \phi_{n}\rightarrow (-1)^n \phi_{n}$ and $\lambda\rightarrow -\lambda$. The onset of long-range couplings in the fractional case ($s<1$) breaks this symmetry, which is restored `little by little' as $s$ approaches unity.

The spatial extent of the localized mode can be measured via the participation ratio $R$, defined as 
\be
R = {(\sum_{n} |\phi_{n}|^2)^2\over{\sum_{n} |\phi_{n}|^4}}
\ee
and is shown in Fig.5 which displays $R$ vs $s$ for several impurity strength values.
Small (large) $R$ values indicate mode localization (delocalization). The most prominent feature of these plots is the observation that the participation ratio decreases with a decrease in the fractional exponent $s$. That is, the mode becomes more and more localized as farthest $s$ is from unity. For a fixed $s$, the participation ratio $R$ also decreases with increasing impurity strength magnitude $|\Delta|$. In all cases, 
$R\rightarrow 1$ when $s\rightarrow 0$. That is, a complete localization of the mode. This can be traced back to 
the dispersion (\ref{eq11}) near $s=0$: $\Omega^2(k)\approx 1+\omega^2+A(k)\;s$, where $A(k)=0.693+\log(1-\cos(k))$. Thus, $\Omega^2(k)\sim 1+\omega^2$ at $s=0$. This means $G_{n 0}(z)=\delta_{n 0}/(z-\omega^2)$ and $G'_{0 0}(z)=-1/(z-\omega^2)^2$. Inserting all this into Eq.(\ref{phin2}), we obtain
$|\phi_{n}|^2 = \delta_{n 0}$. That is, $R=1$.
 
The Green function formalism is also useful to compute the transmission of plane waves across the impurity. 
\begin{figure}[b]
 \includegraphics[scale=0.29]{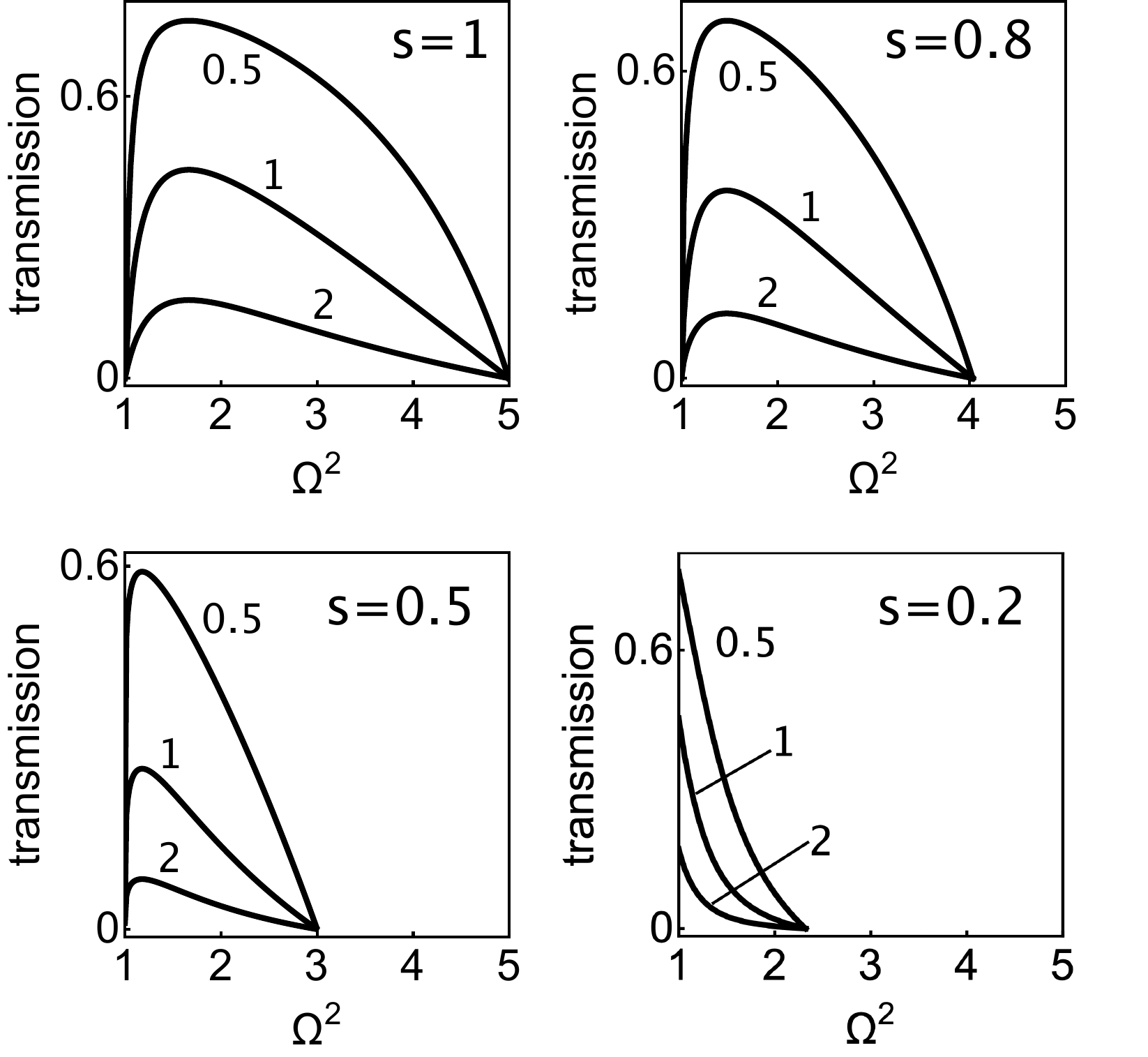}
  \caption{Transmission coefficient across the impurity for different exponents $s$.
  The numbers on each curve denote the impurity strength $\Delta$. ($\omega^2=1$)}  \label{fig6}
\end{figure}
From the general formalism\cite{diecinueve}, the transmission amplitude $T\sim 1/|1-\gamma G_{0 0}^{+}(z)|^2$, while the reflection amplitude $R\sim \gamma^2 |G_{0,0}^{+}(z)|^2/|1-\gamma G_{0 0}^{+}(z)|^2$, where, in our case $\gamma=\Delta(z+\omega^2)$. Let us define a normalization factor $N= T + R$; then the transmission coefficient $t=T/N$ and the reflection coefficient $r=R/N$ will be given by
\begin{eqnarray}
t(z) &=& {1\over{1 + \gamma^2 |G_{0 0}^{+}(z)|^2}}\label{t}\\
r(z)&=&{\gamma^2 |G_{0 0}^{+}(z)|^2\over{1 + \gamma^2 |G_{0 0}^{+}(z)|^2}}, \label{r}
\end{eqnarray}
where $G_{0 0}^{+}(z) = \lim_{\eta\rightarrow 0} \,G_{0 0}(z+i \eta)$ and $z$ is inside the band $\lambda(k)$.  Clearly, $t + r =1$. Also, $t$ (and $r$) is invariant under $\Delta\rightarrow -\Delta$, independent of $s$.

Figure 6 shows the transmission coefficient as a function of the wave frequency $\Omega^2$, for several fractional exponent values and several impurity strengths.
We see the width of the transmission band shrinks with decreasing $s$, while the transmission decreases with increasing $\Delta$. Note that, as $s$ is decreased, 
the left band edge remains fixed but the right edge moves towards the left edge, converging to $\Omega^2=\omega^2$ at $s=0$.

{\em Conclusions}.\ We have examined the effect of using a fractional definition of the Laplacian for a one- dimensional array of coupled electrical units containing a capacitive impurity. The presence of fractionality gives rise to a nonlocal coupling between the electrical units, which at long distances, decreases as a power law, instead of the familiar exponential decrease. In the presence of fractionality, the stationary modes are electrical waves whose dispersion relation was computed in closed form in terms of hypergeometric functions. The main effect of a fractional exponent is the reduction of the bandwidth with decreasing exponent. The localized impurity mode and the transmission coefficient across the impurity were calculated in closed form, using lattice Green functions. The impurity profiles for $\Delta<0$ show a diluted form of the staggered-unstaggered symmetry that one finds in non-fractional ($s=1$) systems. The impurity mode increases its localization with a decrease in fractional exponent $s$. Localization also increases with an increase in impurity strength $\Delta$. The transmission coefficient is not all that dissimilar from the non-fractional case and, as expected, decreases with an increase in $\Delta$. All in all, it seems that the main effect of fractionality is a reduction of the bandwidth as $s$ recedes from the non-fractional case.

Although it might seem that a possible experimental observation of fractionlity would be hard to find in macroscopic systems such as our electrical transmission line, perhaps that could be possible in nanoscopic circuits\cite{cir1,cir2}. The hope is that a nonlocal coupling among units could be emulated taking advantage of the already present $1/|{\bf r}|^3$ decrease of dipole-dipole coupling present in the non-fractional case.

\acknowledgments
This work was supported by Fondecyt Grant 1200120.

\end{document}